# Chirality of nanophotonic waveguide with embedded quantum emitter for unidirectional spin transfer

R. J. Coles[1], D. M. Price[1], J. E. Dixon[1], B. Royall[1], E. Clarke[2], A. M. Fox[1], P. Kok[1], M. S. Skolnick[1] and M. N. Makhonin[1,*]

[1]Department of Physics and Astronomy, University of Sheffield, Sheffield S3 7RH, UK

[2]EPSRC National Centre for III-V Technologies, University of Sheffield, Sheffield S1 3JD, UK

*Correspondence to: m.makhonin@sheffield.ac.uk

**Scalable quantum technologies require faithful conversion between matter qubits storing the quantum information and photonic qubits carrying the information in integrated circuits and waveguides. We demonstrate that the electromagnetic field chirality which arises in nanophotonic waveguides leads to unidirectional emission from an embedded quantum dot quantum emitter, with resultant in-plane transfer of matter-qubit (spin) information. The chiral behavior occurs despite the non-chiral geometry and material of the waveguides. Using dot registration techniques we achieve a quantum emitter deterministically positioned at a chiral point and realize spin-path conversion by design. We measure and compare the phenomena in single mode nanobeam and photonic crystal waveguides. The former is much more tolerant to dot position, exhibits experimental spin-path readout as high as 95 ± 5% and has potential to serve as the basis of future spin-logic and network implementations.**

Quantum information processing promises a dramatically increased performance in computing, secure communications and simulations of quantum systems. In addition, a network of quantum nodes enables distributed quantum computing (*1*), and may facilitate the quantum internet (*2*). Such a distributed architecture requires the conversion between matter qubits for local quantum memories and photonic qubits for quantum communication between nodes (*3, 4, 5*). Furthermore, any scalable quantum technology must solve the miniaturization and fabrication problem, which likely demands that the quantum nodes must be implemented on integrated circuits and waveguides (*6*). Recent developments of passive components (*7*) with embedded quantum dots using advanced semiconductor technologies and enhanced-coherence using resonant techniques on-chip (*8*) contribute to potential solutions. The spin of an electron or hole in a quantum dot (QD) is a promising candidate to serve as the matter-qubit in a quantum network, but its implementation requires an efficient on-chip spin-photon interface. In this work we demonstrate unidirectional transfer of the spin of an embedded dot in a nanophotonic waveguide to a single photon state propagating with directionality factor approaching 100% (95 ± 5%). The results establish a route towards the transfer and entanglement of the spin state of an emitter to a photonic qubit in an integrated network.

Unidirectional light propagation in nanophotonic structures has been reported recently for circularly polarized laser light coupled to surface plasmons (*9*), or scattered into dielectric waveguides by evanescently coupled dipoles (atoms (*10, 11*), gold nanoparticles (*12*)). Unidirectional behavior has also been shown in simulations for photonic crystal waveguides (*13, 14*) and in unpublished work experimentally (*15*). We show here the unidirectionality is general behavior and report it for the important case of an embedded quantum emitter in a nanobeam waveguide, with resultant transfer of spin information. We explain that the spin readout is enabled by the combination of key features of the nanophotonic system: the chiral symmetry of the electromagnetic field, the exact position of the emitter and its spin state, leading to spin-path conversion. We confirm this by QD registration techniques (*16, 17, 18*) to position the dot deterministically at a chiral point and hence achieve highly directional emission by design. We note that our results provide a natural explanation for the bi-directional emission reported but not explained in cross-waveguide structures (*19*).

The system we investigate is a quantum dot in a nanophotonic waveguide. The dot serves as an integrated quantum emitter with addressable spin-exciton eigenstates, generating spin- and position-dependent directional emission of single photons in the waveguide by exciton recombination. Nanophotonic structures support both longitudinal and transverse field components ($E_x$ and $E_y$) and, as we show, permit the transfer of in-plane circular polarization ($E_x \pm iE_y$) (see Fig. 1A for definition of axes). Despite its non-chiral geometry and material, the waveguide exhibits a chiral electromagnetic field mode. As a result, a well-positioned quantum dot will emit in-plane circularly polarized single photons unidirectionally, with a high correlation between the spin state of the quantum dot and the which-path information of the photon.

A schematic of the experimental geometry is given in Fig. 1A (*18*). The left- and right- circularly polarized photons arising from recombination of up and down exciton spin states from a QD embedded within a suspended nanobeam waveguide (NWG) are coupled to NWG modes, and then diffracted by out-coupler gratings at opposite ends towards external detectors. In a non-chiral structure, the emission probability for a circularly polarized dipole is identical in both directions, and this behavior is indeed observed when the dot is located at the center of the waveguide. (See Fig. 1B obtained from FDTD simulations). By contrast, when the dot is displaced from the center of the waveguide, the emission direction depends on the spin, with $\sigma^+$ photons emitted in one direction, and $\sigma^-$ in the other. The unidirectional emission is shown in Figs. 1C and 1D (the results of FDTD simulations), and occurs through the coupling of the $\sigma^+/\sigma^-$ dipole emitters to direction-dependent modes at the chiral points of the NWG.

The origin of the unidirectional emission can be understood by consideration of the electromagnetic field distribution within the laterally-confined nanophotonic geometry. Figure 2A shows a schematic representation of an unterminated nanobeam waveguide with the field distribution in the *xy*-plane at *z = 0* (position of the QDs layer) calculated by solving Maxwell's equations (*18*). The amplitudes of the $E_y$ and $E_x$ field components, together with their relative phases, are shown in Fig. 2B. The $|E_y|$ component has a maximum at the center of the NWG, whereas $|E_x|$ has two maxima closer to the edges (*18*). The relative phase between $E_x$ and $E_y$ is constant at either $\pm\pi/2$ and changes sign when crossing *y = 0*. It is this asymmetry in the phase that enables chiral behavior: the field is right- or left-circularly polarized (i.e. $\sigma^\pm$ corresponding to $E_x \pm iE_y$ fields) at points where $|E_y| = |E_x|$ and the phase is $\pm\pi/2$. This is shown more clearly in Fig. 2C, where the electric field is plotted vectorially at an instant in time. The chiral points are positions where the electric field rotates in time during propagation of the waveguide mode. The rotation direction depends on the propagation direction as indicated in Fig. 2D, so that $\sigma^\pm$ emitters couple preferentially to modes propagating in opposite directions. By contrast, the point at the center of the waveguide (*y = 0*) has no longitudinal component ($E_x = 0$) and is therefore linearly polarized (Fig. 2, C and D). Translation along *y* from the center to the chiral points thus transforms the field polarization from linear, through elliptical, to circular. Note that the propagating electromagnetic field pattern in the waveguide has the characteristics of an azimuthal spin vortex (*20*), where the coupling between spin (polarization) and orbital angular momentum in the waveguide (*12*) leads to rotation of the electric field vector around the vortex core. The relevance of such propagating spin vortices to information processing schemes is discussed in (*21*).

In order to demonstrate the chiral effects experimentally we use a single-mode NWG with randomly distributed self-assembled QDs, positioned by growth in the *xy*-plane at $z = 0$. The guided photoluminescence of single QDs is detected from out-couplers at opposite ends of the NWG (Fig. 1A). The inset in Fig. 3C shows a typical autocorrelation function $g^2(\tau)$ confirming the single-photon nature of the QD emission. The sample was placed in an out-of-*xy*-plane magnetic field $B_z$ to quantize the QD spins into $S_z = \pm 1$ states (spin up/down). This enables identification of exciton Zeeman spin components from the energy of the circularly polarized ($\sigma^+$ / $\sigma^-$) photons emitted: $h\nu_X(S_z) = h\nu_0 + \mu_B g_X B_Z S_Z$, where $g_X$ is an excitonic g-factor that varies from dot to dot with a typical value of ~1.2. $B_Z = 1$ T for the data in Fig. 3. Self-assembled growth leads to random positions of QDs, enabling the observation of emission from both chiral and non-chiral areas of the waveguide. Figure 3 contrasts the behavior for two dots selected from the random distribution, with one showing chiral behavior (Fig. 3A), and the other not (Fig. 3B). For the chiral dot in Fig. 3A, both Zeeman-split components are seen when detecting directly from the dot ("det QD"), but only one component is observed for detection from the left and right out-couplers ("det L", "det R"). The two Zeeman components emit in opposite directions, implying spin-dependent directional emission. By contrast, non-directional emission is observed for the non-chiral dot (Fig. 3B), with both Zeeman components detected from both out-couplers. The degree of contrast in spin-readout (the directionality factor) is defined as:

$$C_{LEFT/RIGHT} = \frac{I_{\sigma+}^{L/R} - I_{\sigma-}^{L/R}}{I_{\sigma+}^{L/R} + I_{\sigma-}^{L/R}},$$

where the superscripts L and R refer to the left and right out-couplers. The results for 50 QDs are shown in Fig. 3C. The diagonal line across the figure is the expectation for a random distribution of dots in the *xy*-plane at $z = 0$ whose circular dipoles couple to the confined electromagnetic fields. QDs at the linear point at the center of the waveguide correspond to zero contrast, whereas dots at the chiral points give rise to contrasts of ±1. Agreement between the trends in contrast between experiment and simulation is seen (the experimental scatter around the diagonal is discussed below).

The data from Fig. 3C are plotted as a histogram of numbers of dots of absolute contrast $C = (|C_{LEFT}|+|C_{RIGHT}|)/2$ in Fig. 3D and compared to binned data of the FDTD simulations of contrast (Fig. S7) (*18*) for out-coupler-terminated waveguides in Fig. 3E (*22*). Qualitative agreement between experiment and simulation is observed, with an increasing number of QDs to high contrast clearly seen, and leveling-off around a contrast of 80%. It is also notable that QDs with extremely high directionality exceeding 90% are observed experimentally (Fig. 3, C and D). These results support the findings in the simulations which show chiral areas (Fig. S6D) of 14% of the waveguide where the contrast exceeds 90% (and 34% exceeding 80%).

The circular dipole unidirectional coupling efficiency, defined as the fraction of power emitted into a waveguide mode propagating in one direction, is calculated to be 68% for a dipole positioned at a chiral point of the NWG (*18*). This is an increase of more than ~2.8 compared to the coupling of a circular-dipole at the center of the waveguide into one spatial direction. It is also ~1.4 times greater than for a linear-dipole co-polarized with the field at the center of the NWG ($y = 0$) (*8*). Both comparisons show the favorable degrees of coupling efficiency obtained for spin readout at the chiral points.

In order to compare to photonic crystal waveguides (PhC WGs), we perform similar experiments and simulations for them. QD spectra with high/low contrasts are shown in Fig. 3F, 3G. The contrasts for the 35 QDs studied are shown in Fig. 3H. The overall trends in the behavior are similar to those seen in Fig. 3C for the NWGs, but with the very marked difference that no QDs exhibit high directionality values above 70% (Fig. 3H, 3J). This is expected as the chiral areas (Fig. S9D) inside the PhC WG where the contrast exceeds 90% only constitute ~0.8% of the waveguide area (~1.5% of area for >80% contrast) (*18*). They are furthermore located in areas of low field intensity (Fig. S9A), leading to poor dipole coupling. Both the high fidelity chiral areas (~20 times smaller than for the NWGs) and field intensities for the PhCs indicate the advantageous characteristics of the NWGs for spin readout, with their near continuous translational symmetry, relative to PhC WGs. This situation can likely be improved for PhC WGs by the use of glide plane techniques, as proposed in (*14*).

As noted above, the experimental points in Fig. 3C and 3H show scatter around the diagonal, compared to theoretical expectations. This behavior is not fully understood. It may arise due to back reflections in the finite length waveguides, in combination with elliptically polarized QDs. Fine structure splitting (*23*) of the quantum dots can be excluded since we observe no magnetic field dependence for asymmetrically coupled QDs. The circular polarization for the QDs may also be affected in photonic structures by the nanostructure and surface proximity (*24*).

Having identified the location of chiral points theoretically, and observed high contrast unidirectional emission for randomly distributed dots, we now demonstrate control of the emission direction, using QD registration techniques (*16, 17, 18*). Firstly, QD spectra were compared before and after the registration and NWG fabrication to confirm that the QD is successfully integrated inside the NWG (Fig. 4A). The waveguide is fabricated (*18*) such that the QD is displaced from the center of the waveguide at the chiral point, with $y_{QD} \approx +90$ nm (Fig. 4B). For this position we expect the $\sigma^+$ dipole to couple to the right propagating mode and the $\sigma^-$ dipole to the left (Fig. 4B). The spectra in magnetic field reveal high directionality ($|C_{LEFT}| = 92\pm3\%$, $|C_{RIGHT}| = 80\pm3\%$) of emission with absolute contrast independent of magnetic field to within 10%. Furthermore, the expected signs of the contrast correlate with the QD position, with the $\sigma^+$ line appearing at the right out-coupler with high energy when the field is positive, while the $\sigma^-$ line is collected from the left out-coupler. Moreover when the direction of the magnetic field is changed, the order of the lines is reversed, but the dipole coupling direction remains the same (Fig. 4C). Thus we not only achieve control of spin-directionality by registration but also the reversal of the emission direction at a given energy by magnetic field control (*15*). We note that the approach with dot registration is not limited to one QD and can be scaled up to create more complex circuits with deterministically coupled QDs to realize spin-photon and spin-spin entanglement on chip.

To conclude, we demonstrate experimentally and theoretically chiral effects in a simple nanophotonic system consisting of waveguides containing embedded quantum emitters. The unidirectional phenomena we report may be used for spin read out and to transfer spin information from localized emitters in waveguide geometries. Deterministic positioning of a QD at a chiral point of the waveguide is achieved, with accompanying spin readout, a possible route to scalability. Larger areas for chiral behavior found in the nanobeam waveguides due to their continuous translational symmetry together with simplified fabrication techniques and low loss may provide significant advantages over photonic crystal waveguides. The findings and techniques presented could pave the way to the creation of novel spin-optical on-chip networks and processing devices based on nanophotonic waveguides.

**Acknowledgments:**

This work has been supported by the Engineering and Physical Sciences Research Council (Programme Grant EP/J007544/1). We thank D. M. Whittaker, D. N. Krizhanovksii, E. Cancellieri and F. Li for fruitful discussions.

**Figure 1**

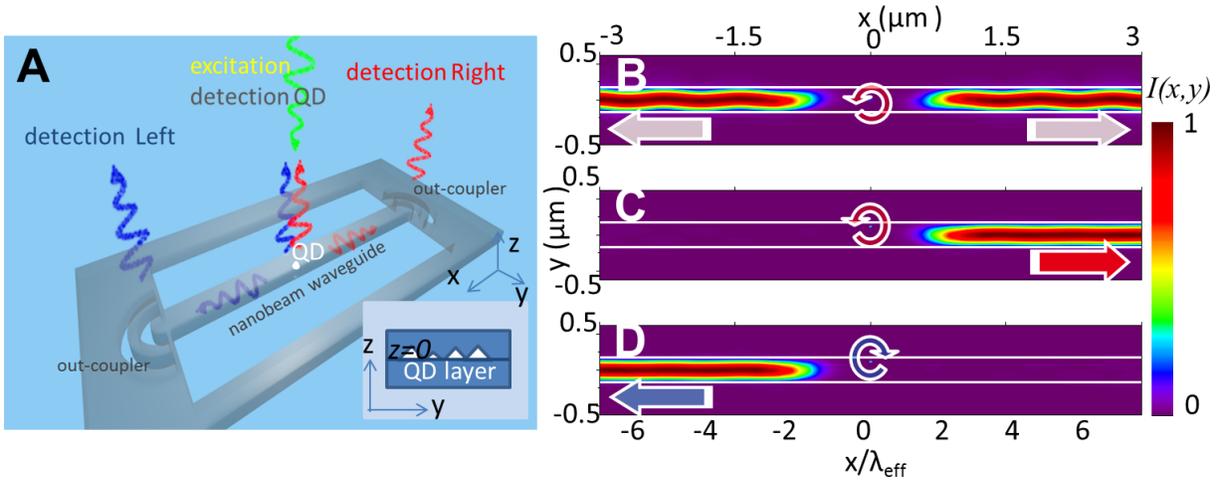

**Fig. 1. Directional emission in dielectric nanobeam waveguides (NWGs). (A)** Schematic representation of the experiments on directional emission from quantum dots embedded inside a single-mode NWG with out-couplers for photon collection. Red (blue) arrows correspond to photons originating from right (left) circularly polarized dipoles in the quantum dot. The inset shows a *yz*-cross section of the NWG with the QD layer at *z* = 0. **(B)** Time averaged intensity distribution of the emission *I(x,y)* in the *xy*-plane at *z* = 0 from circularly polarized $\sigma^+$ dipole at the center of the NWG ( *y* = 0). **(C) & (D)** Emission for a circularly polarized dipole at the chiral point (*y* = *90* nm): **(C)** $\sigma^+$ dipole; **(D)** $\sigma^-$ dipole. The undulation in (B) arises from periodic coupling with time of the rotating circular dipole at the non-chiral point. The intensity distributions in (B-D) are calculated after the dipole source is switched off. The x axis is normalized to the NWG mode effective wavelength $\lambda_{eff}$. The white horizontal lines show the nanostructure boundaries.

**Figure 2**

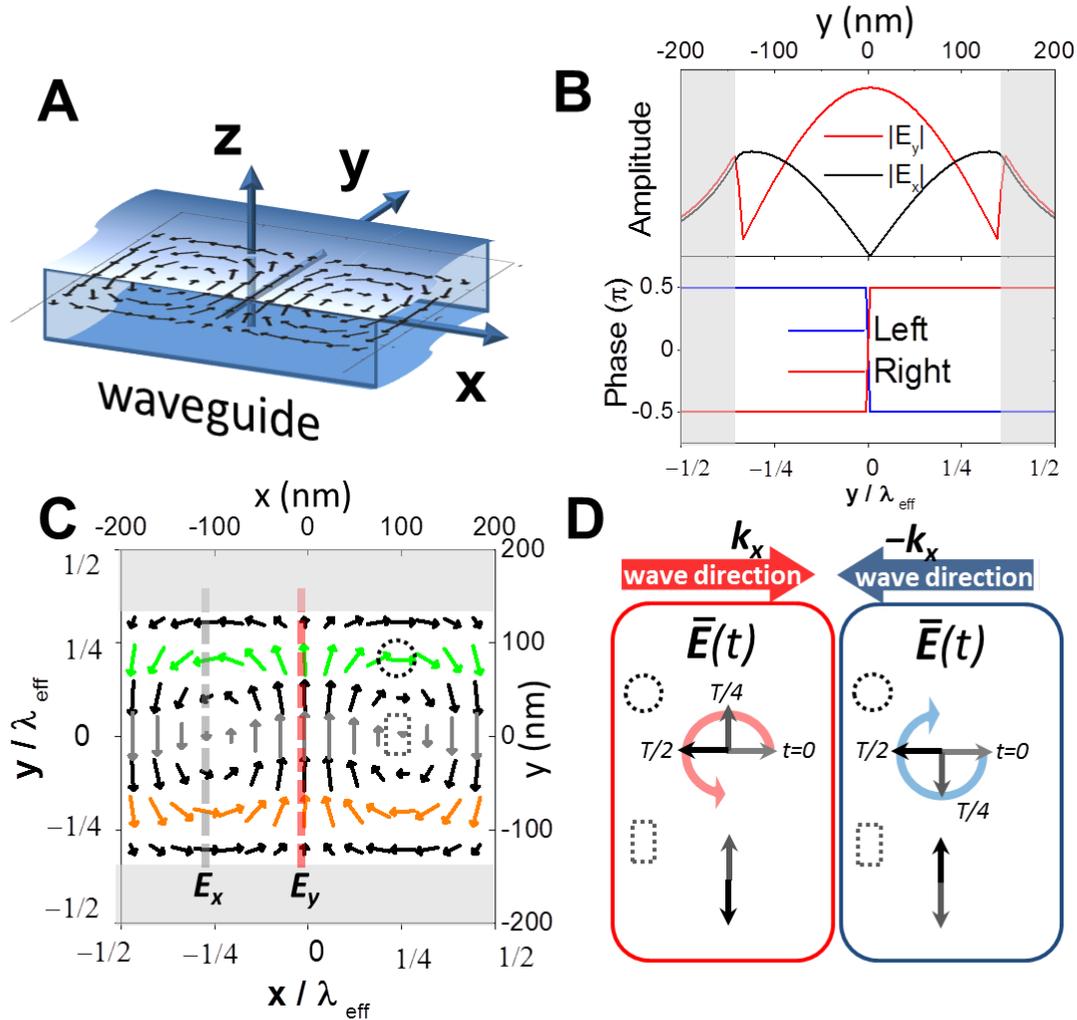

**Fig. 2. Electric field distributions in nanobeam waveguide.** (**A**) Schematic of an infinite length NWG showing the orientation of the axes and the field distribution in the *xy*-plane at *z = 0*, as defined by the central layer of the waveguide containing QDs. (**B**) Electric field amplitudes (top) for the *x* and *y* components and their relative phase (bottom) for left/right propagating modes. (**C**) Simulated distribution of the electric field vector for the first fundamental mode at a fixed moment in time. Arrows show the direction of the field and the length the magnitude. Color is used as a guide to the eye: green/orange for opposite-helicity chiral points, and grey for central points with linear polarization. The left/bottom scales are normalized to the NWG mode effective wavelength $\lambda_{eff}$. (**D**) Electric field time evolution in space from (C) for propagating modes at chiral (circular dotted area) and non-chiral (rectangular dotted area) points in the NWG. The red and blue arrows show the $k_x$ and $-k_x$ mode propagation direction, respectively. The rotation of the electric field vectors exhibiting circular right/left polarization is shown by red/blue circular arrows. The grey areas in (B) and (C) correspond to air cladding regions at the NWG edges.

**Figure 3**

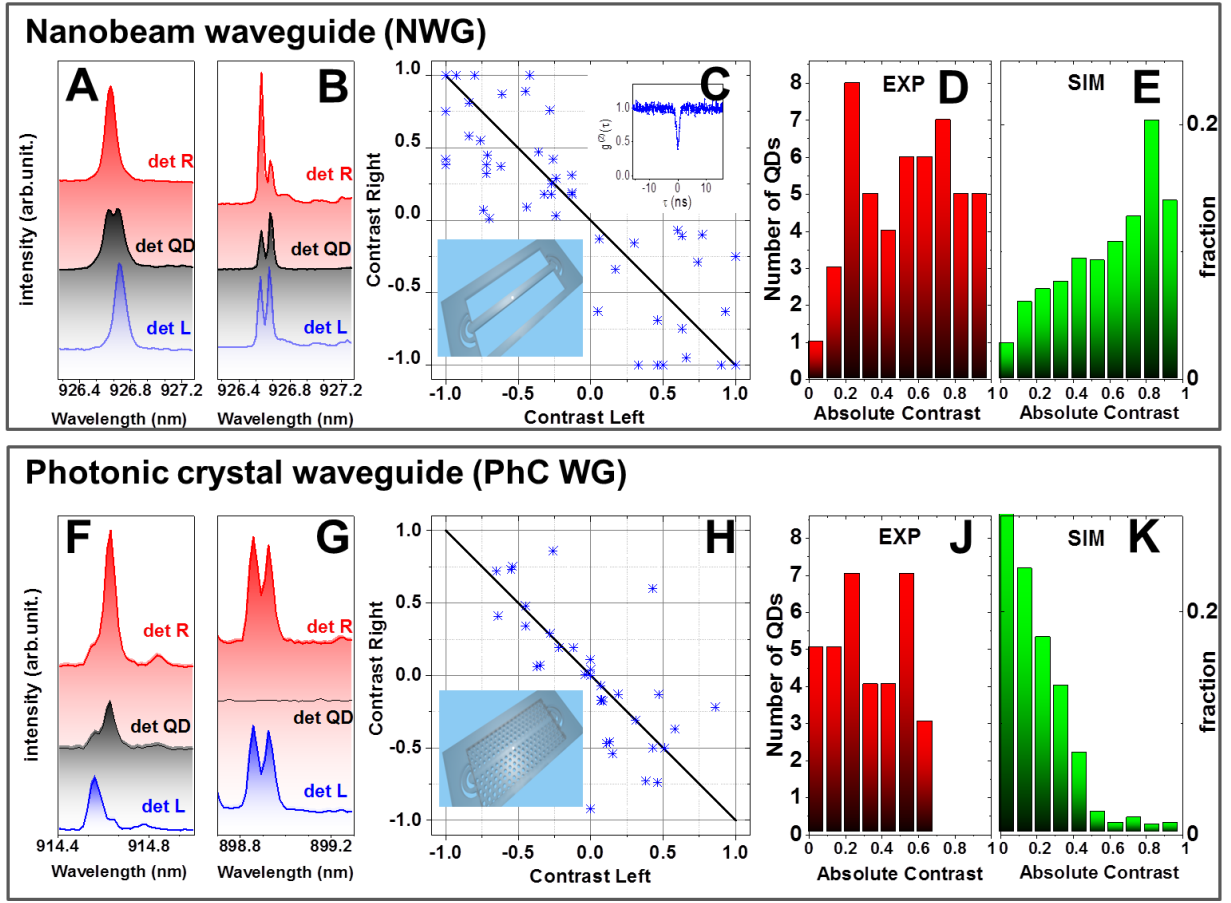

**Fig. 3. Experimental demonstration of spin readout in a NWG and PhC WG.** (**A**) and (**F**) Photoluminescence spectra of QDs with high contrast in NWGs and PhC WGs respectively. (**B**) and (**G**) Experimental photoluminescence spectra of QDs with low contrast in NWGs and PhC WGs respectively. (**C**) and (**H**) Readout contrast $C = (I_{\sigma+} - I_{\sigma-})/(I_{\sigma+} + I_{\sigma-})$ for left and right out-couplers for randomly distributed self-assembled QDs in NWGs and PhC WGs, respectively. The points correspond to experimental data, and the black diagonal lines show the expected distribution for ideal circularly polarized dipoles. (**D**) and (**J**) Statistical distributions of the absolute average contrasts $C = (|C_{LEFT}|+|C_{RIGHT}|)/2$ of QD spin readout taken from the experimental data in (C) and (H) for NWGs and PhC WGs, respectively. All experimental data are recorded at $B_Z = 1$ T. (**E**) and (**K**) Simulated statistical distributions of absolute contrasts calculated from the coupling of circularly-polarized dipoles distributed across the WGs in the *xy*-plane ($z = 0$) taking into account the electric field distributions (*18*) due to the out-couplers for NWG and PhC WGs respectively. The inset in (C) shows a typical QD autocorrelation function $g^2(\tau)$. The insets in (C) and (H) show schematics of the NWG and PhC WGs respectively.

**Figure 4**

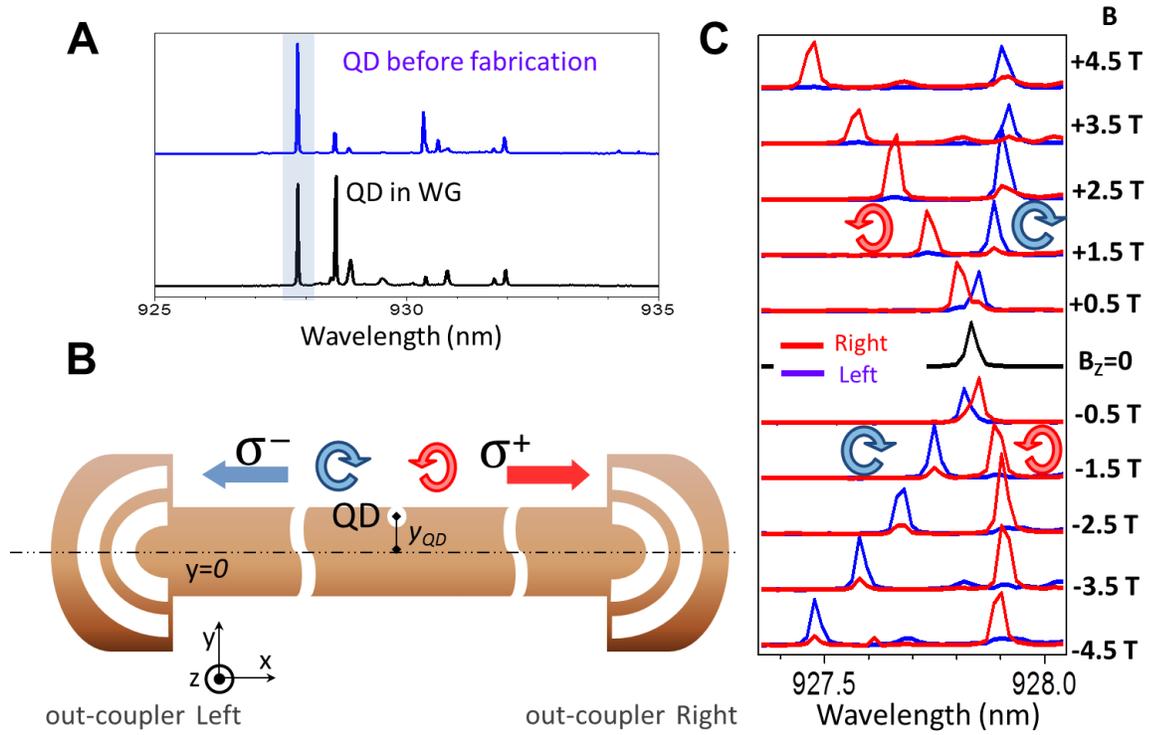

**Fig. 4. Quantum dot registration and spin readout.** (**A**) Spectra of the QD before and after NWG fabrication. (**B**) Schematic of spin readout of a registered QD that is displaced vertically by $y_{QD}$ = *90* nm in a NWG of *280* nm width (*18*). (**C**) Magnetic field dependence of emission from the registered QD collected from the left out-coupler (blue) and right out-coupler (red).

# Supplementary Information for

# Chirality of nanophotonic waveguide with embedded quantum emitter for unidirectional spin transfer

R. J. Coles, D. M. Price, J. E. Dixon, B. Royall, E. Clarke, A. M. Fox, P. Kok, M. S. Skolnick and M. N. Makhonin

**Content:**

Materials and Methods

Supplementary Text

Figs. S1 to S10

# Materials and Methods

*I. Sample Details*

The sample was grown by molecular beam epitaxy on an undoped [100] GaAs substrate, and consisted of a *140* nm thick GaAs membrane containing a single layer of self-assembled InGaAs quantum dots (QDs) at the center grown on a *1* μm sacrificial $Al_{0.6}Ga_{0.4}As$ layer. The photonic structures were patterned by electron beam lithography and the pattern transferred to the membrane using inductively coupled plasma etching. To release the membranes, the sacrificial AlGaAs layer was removed by selective wet etch using a buffered hydrogen fluoride solution. The suspended nanobeam waveguides were *15* μm long, *280* nm wide and *140* nm height. An SEM image of the suspended nanobeam waveguide is shown in Figure S1(a). The W1 photonic crystal waveguides were fabricated with lattice constant $a = 254$ nm and hole radius $r = 0.31\ a$. An SEM image of the photonic crystal waveguide is presented in Figure S1(b). Both waveguides were terminated with semi-circular $\lambda/2n$ air-GaAs out-coupler gratings designed for optimum operation at $\lambda = 950$ nm at the center of the QD ensemble photoluminescence (PL) emission *(25)*.

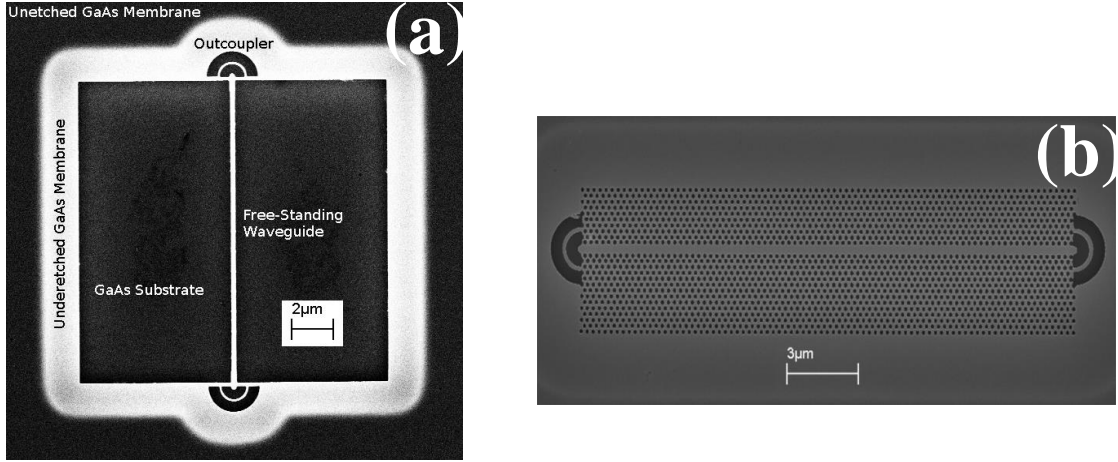

*II. Experimental Setup*

Figure S1. Example SEM images of (a) suspended nanobeam waveguide and (b) photonic crystal waveguide used in experiments.

All measurements were performed in a helium bath cryostat at $T = 4.2$ K within which a superconducting magnet provided magnetic field of up to *5* T normal to the sample plane. Ultra-stable positioning of the sample within the system is provided by X, Y, Z home-made piezo stages. The cryostat insert had optical access to the sample in a confocal scanning microscope arrangement *(26)*. The excitation and collection spots were below *1.5* μm in diameter and could be separately moved by more than *15* μm by scanning mirrors to obtain the exact geometry required for each experiment. Optical excitation was provided by a *808* nm diode laser coupled to an optical fiber. The collected PL signal was fiber coupled and spectral measurements were performed using a spectrometer with a LN2 cooled CCD. Using a second exit port on the spectrometer, spectrally filtered PL signals were sent to two avalanche photodiodes (APDs) for QD autocorrelation $g^2(\tau)$ measurements using a single photon counting module.

## III. Quantum Dot Registration Method

Dot registration *(16, 17)* is carried out in a scanning micro-photoluminescence (µPL) set-up with two collection paths, as illustrated in Figure S2. The dot registration process involves three stages: pre-registration markers are fabricated, individual quantum dots are registered, and photonic structures are deterministically fabricated around a quantum dot.

Registration markers are patterned using electron beam lithography (EBL) and a positive resist. The sample is developed in Xylene leaving the wafer surface exposed in the desired pattern, as shown in Figure S3 (a). The inner (green) markers are used to register the relative position of a quantum dot, whilst the outer (blue) markers are used to re-align the EBL during the final fabrication stage. A Layer of *5* nm of Titanium followed by *20* nm of Gold are then evaporated onto the wafer. The sample is then placed in a solvent bath to remove the remaining layer of resist and any unwanted Titanium and Gold.

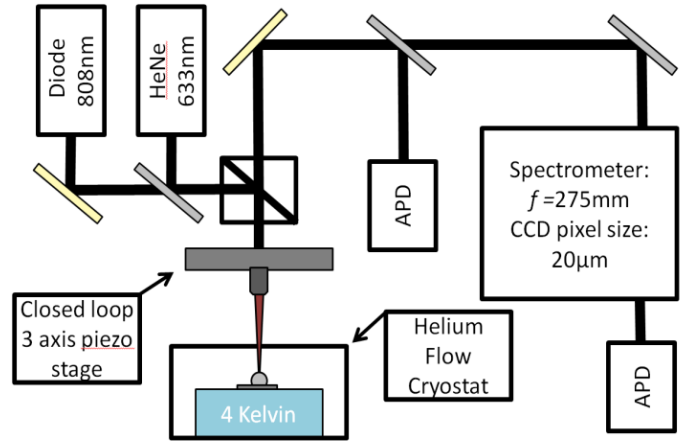

Figure S2. Experimental setup used for dot registration. An *808* nm diode laser is used to excite the sample through a *50:50* cubic beam splitter. A 50x objective with NA of *0.5* is used to focus on to the sample. The collected light is split by a *92:8* pellicle beam splitter (Transmission: Reflection) and is detected via two independent APDs. The sample is mounted in an Oxford Instruments continuous flow cryostat and cooled to *4* K.

A *1* mm diameter cubic zirconia ($ZrO_2$) Weierstrass solid immersion lens (SIL) is placed directly above a registration grid, with an aberration free image size of ~ *50* µm in diameter. A low quantum dot density of ~ $1.6 \cdot 10^7$ cm$^{-2}$ and a SIL-enhanced excitation spot size ~ *350* nm enables individual quantum dots to be measured. The closed-loop 3 axis scanning piezo stage, with a resolution of *1* nm, is used to scan the objective lens which simultaneously moves the excitation and collection spots. One collection path is spectrally filtered through a monochromator to isolate a single exciton line before it is measured with an APD. The second collection path is used to measure the reflected laser signal from the gold markers to provide reference points in the scan from which the QD position is measured. No additional spectral filtering is used on the second collection path before it is detected with an APD, although it is heavily attenuated using neutral density filters. Horizontal and vertical line scans are used to determine the relative quantum dot position. The signals from both APDs were recorded during the scans as a function of the stage position (Figure S3 (b)).

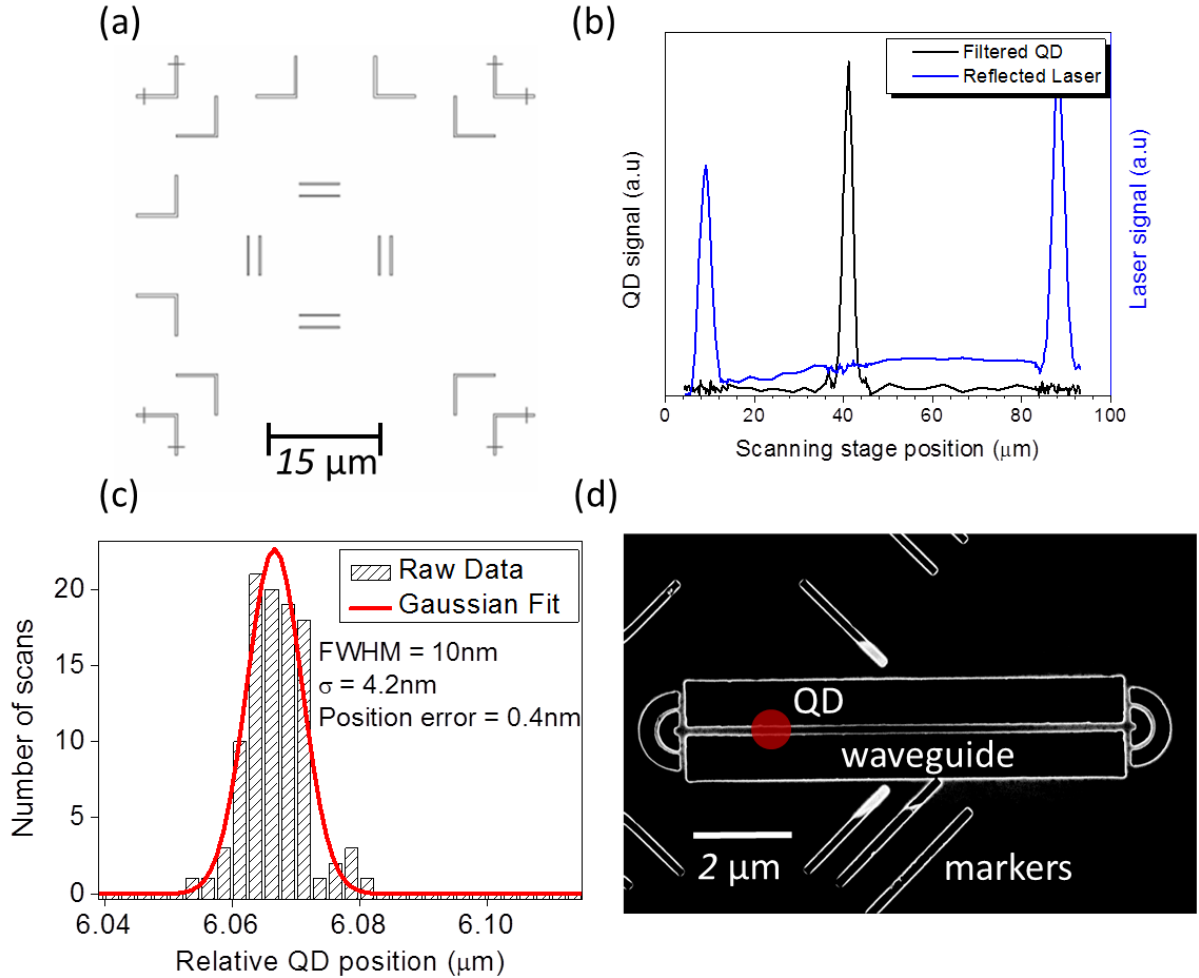

Figure S3. (a) Schematic showing design of gold markers on the surface of the sample. (b) Example line scan showing how the vertical position is determined relative to a gold marker. The bright laser peaks are used to determine the magnification of the system due to the SIL (typically ~5.4). (c) Histogram of relative quantum dot positions measured for a single dot in the vertical axis. An error of 0.4 nm is calculated as $\sigma/\sqrt{N}$ where σ is the standard deviation and N is the total number of scans. (d) SEM of the fabricated waveguide with QD position marked with a red circle.

The spacing between the gold markers is designed to be *15* μm. The relative position of the quantum dot is then determined by repeating the line scans 100 times to provide a statistical average (Figure S3(c)). A Gaussian curve is fitted to the distribution to determine the center. The minimum standard deviation observed is σ < 5 nm which is comparable to the best uncertainty observed in the literature (*16, 17*). The position of the suspended nanobeam waveguides to be fabricated is placed within the original EBL design file at a suitable location, such that the QD are in the center or laterally displaced. To align the EBL writing field to the pattern previously fabricated, the electron beam is scanned over the outer (blue) registration markers. The reflected electron flux is recorded and used to determine the central position of each marker. The suspended nanobeam is then fabricated (Fig. S3 (d)) as described above in Section I.

# Supplementary Text

*IV. Approximate solution for the field profiles of Suspended Nanobeam Waveguides*

An approximate form of the fields of the suspended nanobeam waveguides can be found by making a few simplifying assumptions and solving Maxwell's equations with appropriate boundary conditions. Firstly, the fields are assumed to be of the form $\mathbf{E}(x, y, z, t) = \mathbf{E}(y, z)\,e^{i\omega t + ikx}$, where $k = 2\pi/\lambda$ is the propagation wavenumber of the guided mode with wavelength $\lambda$ and angular frequency $\omega$ (see Fig. 2A of main text for schematic of waveguide and axes). The electric dipole of the quantum dot is constrained to the *xy*-plane so that the only modes of interest are those which have an even parity in *z* such that $\mathbf{E}(z) = \mathbf{E}(-z)$ (*8*). The fields therefore have a cosine-like z-dependence, and at $z = 0$ (the center of the GaAs membrane) $d\mathbf{E}/dz = d\mathbf{H}/dz = 0$. The fields in the $z = 0$ plane can thus be written as:

$$kE_y - i\frac{dE_x}{dy} = \mu\omega H_z, \quad \text{(Eq. 1a)}$$

$$\frac{dH_z}{dy} = i\omega\omega_x \quad \text{(Eq. 1b)}$$

$$kH_z = -\varepsilon\omega E_y. \quad \text{(Eq. 1c)}$$

The fundamental TE-like mode of the suspended nanobeam waveguide has odd parity in *y* so $E_y(y) = E_y(-y)$ and $E_x(y) = -E_x(-y)$. This means that the fields internal to the waveguide are given by:

$$E_y(y, z=0) \propto \cos(\tilde{k}y) \quad \text{(Eq. 2a)}$$

$$E_x(y, z=0) \propto -i\sin(\tilde{k}y) \quad \text{(Eq. 2b)}$$

$$H_z(y, z=0) \propto -\cos(\tilde{k}y) \quad \text{(Eq. 2c)}$$

where $\tilde{k} = k\sqrt{n^2+1}$ and $n = 3.4$ is the refractive index of GaAs. There are important conclusions to be drawn from these equations: i) the only fields in the QD plane are $E_x$, $E_y$ and $H_z$; ii) there is a $|\varphi| = \pi/2$ phase shift between $E_x$ and $E_y$ as shown in Figure 2B of the main text; iii) the sign of $\varphi$ is inverted when either the propagation direction ($\pm k$) or dipole position ($\pm y_d$) change sign. These properties are confirmed by numerical simulations of the suspended nanobeam in three dimensions as presented in the main text in Figure 2B and discussed below.

*V. Simulations of Infinite Suspended Nanobeam Waveguide*

Fully-vectorial eigenmodes of Maxwell's equations with periodic boundary conditions were computed by preconditioned conjugate-gradient minimization of the block Rayleigh quotient in a plane-wave basis, using a freely available software package (*29*). The complex fields of the fundamental propagating mode of a 2D *yz*-cross section of the suspended nanobeam waveguide were calculated for $\lambda = 930$ nm using the dimensions given in Section I, the results of which are shown in Figure 2B of the main text. The degree of readout contrast for left and right propagation was calculated from the modal fields of Figure 2B using Eq. 3(a) and (b) respectively:

$$C_L = \frac{(\Gamma_L^- - \Gamma_L^+)}{(\Gamma_L^- + \Gamma_L^+)},\qquad\text{(Eq. 3a)}$$

$$C_R = \frac{(\Gamma_R^+ - \Gamma_R^-)}{(\Gamma_R^+ + \Gamma_R^-)}.\qquad\text{(Eq. 3b)}$$

The QD emission rate (*28*), $\Gamma \propto Re\{\mathbf{d}^*\cdot\mathbf{E}\}^2$, where the dipole moment is $\mathbf{d}_\pm = d(\hat{\mathbf{e}}_x \pm i\,\hat{\mathbf{e}}_y)$ and the local electric fields are $\mathbf{E} = |E_x|\,\hat{\mathbf{e}}_x + |E_y|\,e^{i\varphi}\,\hat{\mathbf{e}}_y$. The dipole moment $\mathbf{d}_+$ ($\mathbf{d}_-$) denotes the dipole moment of a $\sigma^+$ ($\sigma^-$) circularly polarized dipole source. The calculated contrast for both propagation directions is shown in Figure S4. The fractional area over which contrast exceeding 90% (80%) is observed is 29% (41%) for the infinite nanobeam waveguide.

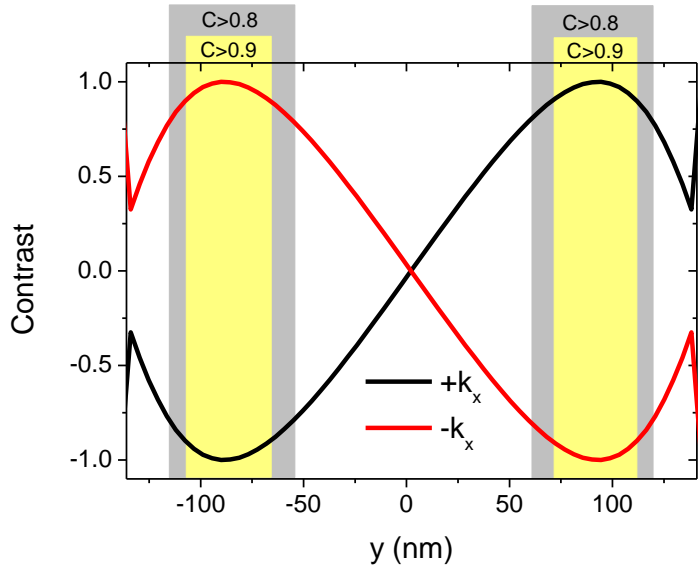

Figure S4. Calculated contrast for infinite length nanobeam waveguide $\mathbf{d}_+$ ($\sigma^+$ polarized) and $\mathbf{d}_-$ dipoles ($\sigma^-$ polarized) for propagation to the left ($-k_x$) is shown in red and to the right ($+k_x$) is shown in black.

## *VI. Simulations of Out-coupler Terminated Suspended Nanobeam Waveguide*

Presentations of simulations in the main text include data calculated for waveguides terminated by out-coupler gratings only in Fig. 3E. The out-coupler designs from (*25*) were used for the experiments presented in this paper so this section investigates the behavior of a suspended nanobeam waveguide terminated by out-couplers of this design. Complete simulated data and analysis for the effects of the out-coupler gratings are considered here. The out-coupler gratings have back reflections. A commercial-grade simulator based on the finite-difference time-domain method was used to perform the calculations for the nanobeam waveguide terminated by grating out-couplers (*29*). The waveguide dimensions were the same as specified above in section *I*. The waveguide mode was excited using a mode source and the field profiles recorded using a field monitor in the *xy*-plane (*z = 0*) centered at *x, y = 0*. Due to the calculated reflectivity of ~15% from the out-coupler gratings, standing waves are formed in the waveguide. The calculated field profiles for the $E_x$ and $E_y$ field components internal to the waveguide are shown in Figures S5(a) and (b) respectively for $\lambda = 934$ nm, corresponding to a Fabry-Perot resonance maximum. The calculated phase distributions for these fields are shown in Figure S5(c) and (d) for excitation from the left and right sides of the waveguide respectively. Due to reflections from the out-couplers, a longitudinal phase variation is introduced where $0.345\pi \leq |\varphi(x)| \leq 0.655\pi$. Regions where the phase is locally $|\varphi(x)| = \pi/2$ are shown by vertical white contours in Figure S5(c) and (d).

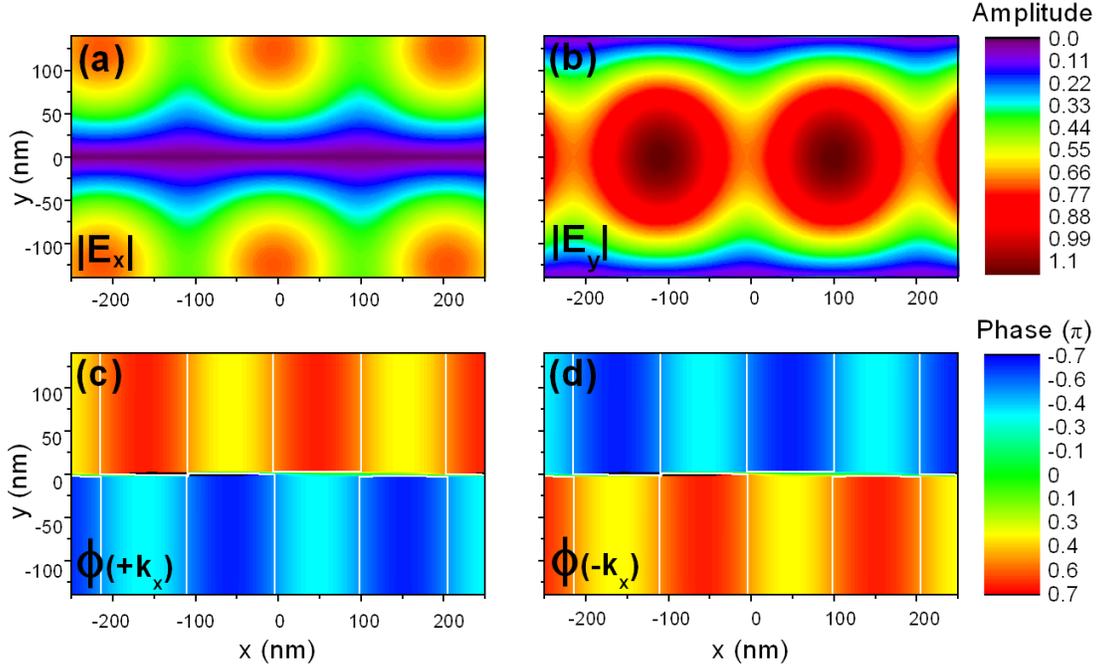

Figure S5. Calculated electric field profiles for (a) $E_x$ and (b) $E_y$ components for the nanobeam waveguide terminated by the out-couplers. Phase between the electric field components for the same waveguide for mode excitation from the (c) left and (d) right. White vertical contours indicate where local field phase $|\varphi(x)| = \pi/2$.

To gain a more intuitive insight into the effects of standing wave formation on the field polarization profile, Stokes parameters are calculated for the electric fields of Figure S5. The Stokes parameters are defined in Eqs. 4(a-d) as:

$$I = |E_x|^2 + |E_y|^2 \qquad \text{(Eq. 4a)}$$

$$Q = |E_x|^2 - |E_y|^2 \qquad \text{(Eq. 4b)}$$

$$U = 2\,\text{Re}\{E_x E_y^*\} \qquad \text{(Eq. 4c)}$$

$$V = -2\,\text{Im}\{E_x E_y^*\} \qquad \text{(Eq. 4d)}$$

The intensity $I$ is shown in Figure S6(a), the degree of linear polarization Q in (b), the degree of diagonal polarization U in (c) and circular polarization V in (d). Regions where the circular polarization degree approaches unity are herein referred to as 'C-points'.

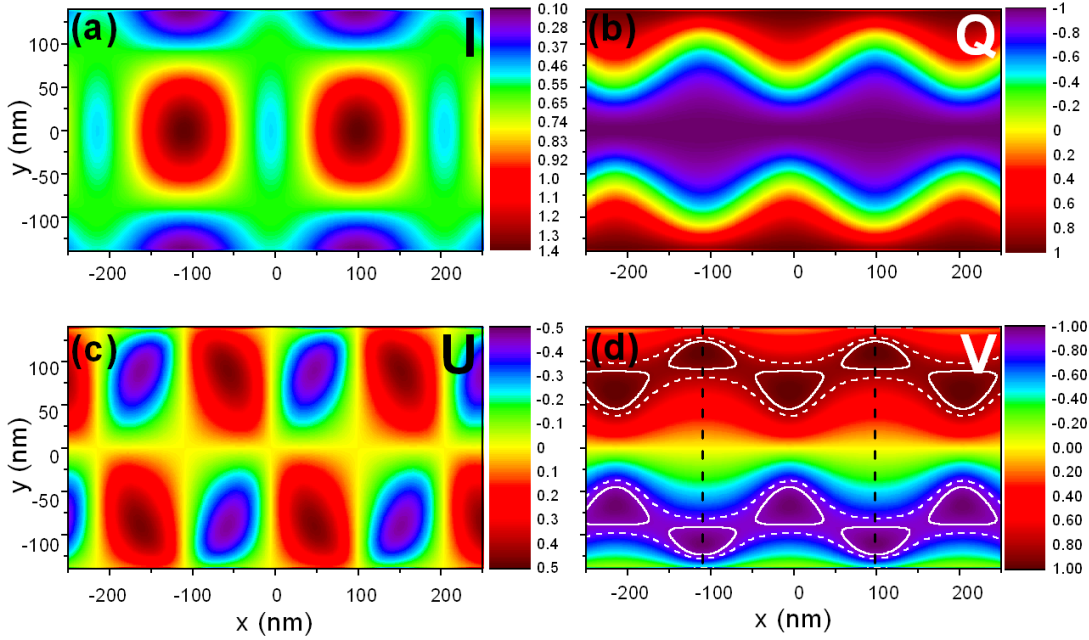

Figure S6. Calculated Stokes parameters for the fields of the nanobeam waveguide terminated by out-couplers. (a) Intensity, *I*. (b) Degree of linear polarisation, *Q*. (c) Degree of diagonal polarisation, *U*. (d) Degree of circular polarisation, *V*. Solid (dashed) white contours enclose regions where the degree of circular polarisation exceeds 90% (80%). Black dashed lines show the 'unit cell' used for statistical calculations in the main text.

As with the infinite length nanobeam waveguide, close to the center of the waveguide the fields are linearly polarized along *y*. The standing waves of the waveguide establish regions of diagonal polarization for *|y|>0* with periodicity along *x*. The regions of high (*|V|>0.9*) circular polarization are also periodic along the waveguide axis. However for *|y| ~ 90* nm the circular polarization degree remains above *0.87* for all values of *x*. At the same displacement, the infinite suspended nanobeam case yields a circular polarization degree of *>0.99* continuously along *x*. The modulation of the mode profiles from back reflections from the out-couplers acts to reduce the maximum degree of circular polarization at a given position in the waveguide. Appropriately designed impedance matched out-couplers can eliminate these back reflections and restore behavior closer to the infinite suspended nanobeam waveguide case.

Calculation of the contrast of spin readout as in the previous section yields spatial profiles which exhibit the same spatial dependence as the degree of circular polarization of the suspended nanobeam waveguide fields, as shown in Figure S7. Despite the effects of the out-couplers, high fidelity (*C > 0.99*) spin readout contrast is observed along the waveguide as demonstrated experimentally in Figure 3D and in simulations presented in Figure 3E of the main text. The fractional area over which contrast exceeding 90% (80%) is observed is 14% (34%) for the nanobeam waveguide terminated by out-coupler gratings. Compared to the infinite waveguide case these areas show a reduction in size of ~ 52% for regions exceeding 90% contrast (~ 17% for regions exceeding 80%) contrast due to back reflections from the out-couplers.

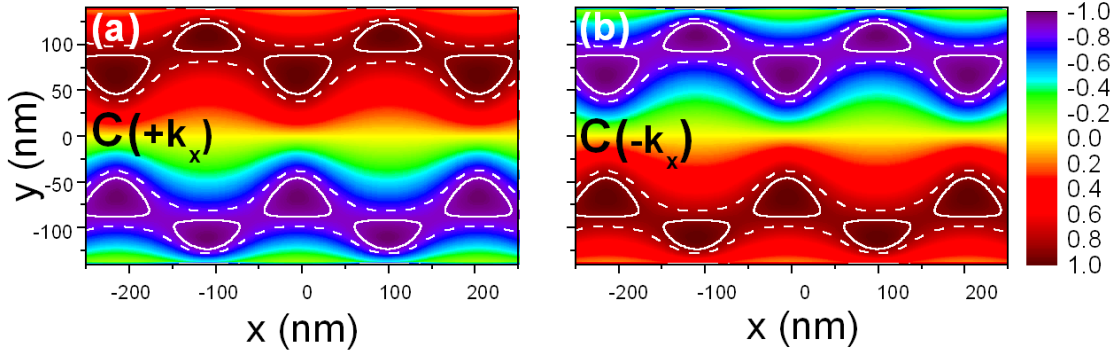

Figure S7. Calculated contrast of spin readout for the nanobeam waveguide terminated by out-couplers for propagation to the (a) right and (b) left. White solid (dashed) contours enclose regions of contrast exceeding 90% (80%).

## *VII. Simulations of Photonic Crystal Waveguide*

Fully-vectorial eigenmodes of Maxwell's equations with periodic boundary conditions were computed for the photonic crystal waveguide using a freely available software package *(29)*. The waveguide dimensions were the same as specified in Section I. The waveguide mode was excited using a mode source and the field profiles recorded using a field monitor in the *xy*-plane ($z = 0$) centered at x, y =0. The Bloch periodicity of the photonic crystal leads to the formation of longitudinal and lateral variations of the field profiles. All figures presented for the photonic crystal are for infinite waveguides. The field profiles of the W1 photonic crystal waveguide mode are shown in Figure S8 (a) and (b) with the phase between them for propagation to the right ($+k_x$) in Figure S8 (c).

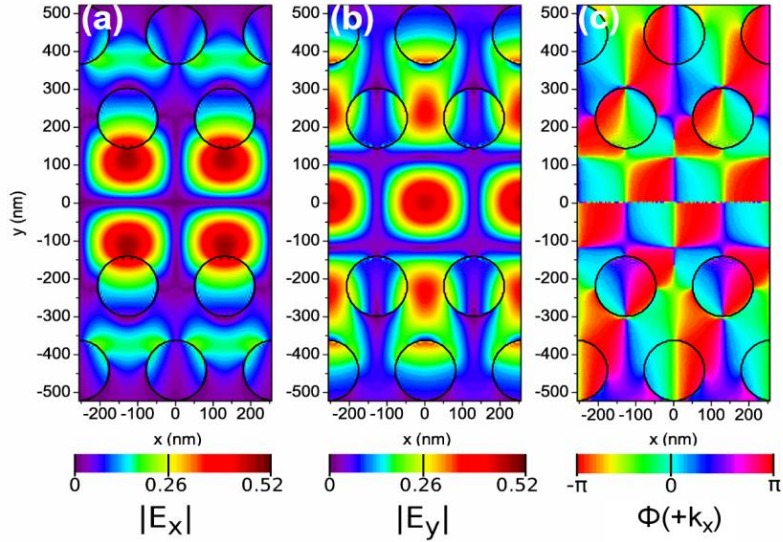

Figure S8. Calculated electric field profiles for (a) $E_x$ and (b) $E_y$ components for the infinite photonic crystal waveguide. Phase between the electric field components for the same waveguide for mode propagation to the right ($+k_x$). For propagation to the left ($-k_x$), the phase is inverted. Black contours enclose air holes of photonic crystal membrane.

The Stokes parameters are calculated for the electric fields of the photonic crystal waveguide mode at $\lambda = 934$ nm. The intensity is shown in Figure S9(a), the degree of linear polarization Q in (b), the degree of diagonal polarization U in (c) and circular polarization V in (d). The strong modulation of the modal fields by the photonic crystal lattice produces regions of fully diagonally polarized fields. There are 6 C-points (C > 90%) per unit cell for the photonic crystal waveguide compared to the quasi-continuous bands of high circular polarization for the suspended nanobeam waveguide with out-couplers. In addition, for the W1 photonic crystal waveguide, these points occur near to regions of low field intensity so a low dipole coupling efficiency is expected.

The spin readout contrast for circularly polarized dipole sources was calculated for the photonic crystal waveguide fields, using the same method as for the suspended nanobeam in the previous section. The results of these calculations are plotted in Figure S10. The calculated spatial distribution of the contrast exhibits the same spatial profile as the degree of circular polarization V from Figure S7(d). This is consistent with experimental results in Figure 3J which demonstrates no QDs with a contrast of spin readout exceeding 0.7. The reduced average fidelity for the photonic crystal waveguide relative to the suspended nanobeam waveguide is a direct consequence of the limited number and size of the chiral points provided by the modal fields of the photonic crystal waveguide. The fractional area over which contrast exceeding 90% (80%) is observed is 0.8% (1.5%) for the photonic crystal waveguide, which is around 5% of the area observed in the suspended nanobeam waveguide with out-couplers.

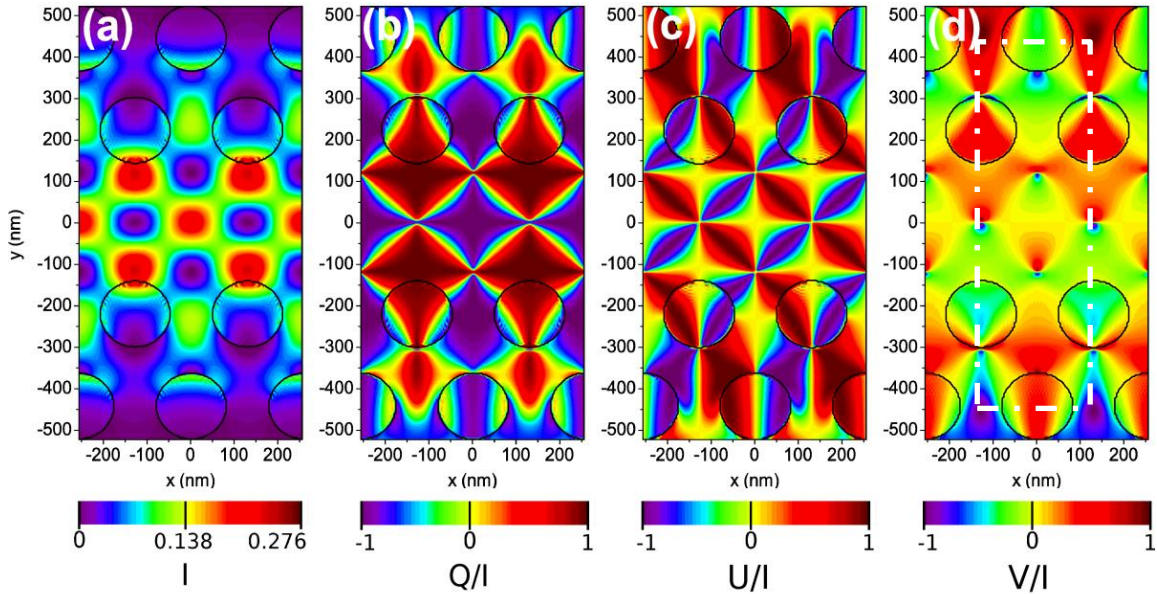

Figure S9. Calculated Stokes parameters for the fields of the photonic crystal waveguide. (a) Intensity, I. (b) Degree of linear polarization, Q. (c) Degree of diagonal polarization, U. (d) Degree of circular polarization, V. Black contours enclose etched (air) regions of the photonic crystal lattice. White dash-dot region shows unit cell used for statistical calculations in the main text. Data in the air holes was omitted from calculations. Regions of high circular polarization degree are not highlighted in (d) as these regions are of small area, but can been seen as points at the extremes of the color scale.

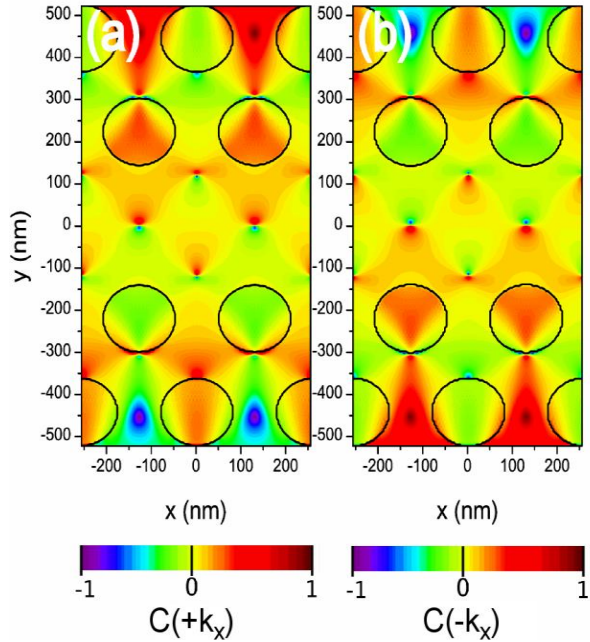

Figure S10. Calculated contrast of spin readout for the photonic crystal waveguide for propagation to the (a) right and (b) left.